\documentstyle[aas2pp4,epsf]{article}
\lefthead{TAKAGI, ARIMOTO \& VANSEVI\v CIUS}
\righthead{Dust in Starburst Galaxies}

\begin{document}

\title{Age and Dust Degeneracy for Starburst Galaxies Solved?}

\author{Toshinobu TAKAGI}
\affil{Department of Physics, Rikkyo University \\
        3-34-1 Nishi-Ikebukuro, Toshima-ku, Tokyo 171-8501, Japan; \\
      }
\author{Nobuo ARIMOTO}
\affil{Institute of Astronomy, School of Science, University of Tokyo \\
        2-21-1, Osawa, Mitaka, Tokyo 181-8588, Japan; \\
      }
\author{Vladas VANSEVI\v CIUS}
\affil{National Astronomical Observatory \\
        2-21-1, Osawa, Mitaka, Tokyo 181-8588, Japan; \\
       Institute of Physics, Go\v stauto 12, Vilnius 2600, Lithuania \\
       }

\begin{abstract}

 A spectral evolution model of galaxies that includes both stellar and 
dust effects is newly built. 
xApplying the model to 22 nearby starburst galaxies,
we have shown that far infrared luminosity of galaxies helps to break the
age-dustiness degeneracy. We have derived a unique solution of
age and the dustiness for each starburst galaxy.
The resulting starburst ages and optical depths are in the
range $10 \le t \ ({\rm Myr}) \le 500$ and $0.5 \le \tau_{V} \le 5.0$,
respectively. The result is robust and is almost independent of model
assumptions such as  
dust distributions, extinction curves, and burst strengths.
With the rapidly growing sensitivity of submillimeter detectors, it
should become possible in the near future
to determine the age and $\tau_{V}$ of 
star-forming galaxies at redshifts $z \simeq 3$ and beyond.
Accurate estimates of $\tau_{V}$ for Lyman-break galaxies 
and high-z galaxies might 
require a substantial revision of the previously claimed picture of star
formation history over the Hubble time.

\end{abstract}


\section {INTRODUCTION}

 Starburst (SB) galaxies in the nearby Universe are the best sites to
study dust. Star formation in the SB galaxies should occur in localized
regions, where massive OB stars are deeply embedded in gas clouds
that contain significant amounts of dust on average. The bulk of the photons
emitted from burst populations are absorbed and partly scattered
by dust, and are re-radiated in at far infrared (FIR) wavelengths.
The intrinsic spectral energy distribution
(SED) of the SB population is thus considerably modified
in the UV. This applies not only to the nearby SB galaxies,
but also to star-forming galaxies at high redshifts, since a timescale
of heavy element synthesis in massive galaxies at an initial
burst of star formation should be as short as $\sim 10^8$ yr
(e.g., Arimoto \& Yoshii 1987), and thus galaxies at high
redshifts should contain as much dust as the nearby galaxies do. 
In recent years, there have been many attempts to derive the history of the 
star formation rate (SFR) in the Universe, e.g., by converting the UV light 
densities to the SFRs of the CFRS (Canada France Redshift Survey)
galaxies at $0 \le z \le 1$ (Lilly et al. 1996) and the HDF 
(Hubble Deep Field) galaxies at $1 \le z \le 4$ (Madau et al. 1996; 
Connolly et al. 1997).
However, without precise estimates of the dust extinction and stellar age,
it could be misleading to use the observed UV flux density
as an indicator of the SFR, because the UV light is most sensitive to dust
obscuration (Meurer et al. 1997; Rowan-Robinson et al. 1997; Calzetti
1998). Unfortunately, the UV-near infrared (NIR) SEDs are 
degenerate in stellar
age and dust attenuation (Gordon, Calzetti, \& Witt 1997). 
Complicated analyses of the SEDs are required to solve this
degeneracy by incorporating the dust effects into a model of galaxy spectral
evolution. The nearby SB galaxies are most suitable to
calibrate the dust effects in actively star forming environments.
Once the degeneracy is solved, an attempt can be made
immediately to determine the ages of Lyman-break galaxies (Steidel et
al. 1996) at high redshifts which are obviously too red to be dust-free
star-forming galaxies (Meurer et al. 1997).
It would also be possible to derive the age of primeval galaxy
candidates in which the bulk of stars of present-day ellipticals formed
in a relatively shortlived starburst for the first time
(Hughes, Dunlop, \& Rawlings 1997).

 Gordon et al. (1997) studied the UV-NIR SEDs of 30 SB
galaxies which manifest star-forming activity by the strong UV SEDs
and intense line emissions. These are irregular and
spiral galaxies with disturbed morphologies 
locating at a median distance 60 Mpc
(hereafter H$_0$=50 km s$^{-1}$ Mpc$^{-1}$). Gordon et al. (1997) 
found that these galaxies show little evidence for
a bump at $0.22 \mu m$, suggesting that the extinction
curve similar to that of the Small Magellanic Cloud
(SMC) is more suitable than that found for
the Milky Way (MW). The best geometry that describes the
SB region was found to have the inner dust-free 
cavity having star clusters surrounded
by the dust shell with local clumpy dust condensations. Gordon et al.
(1997) tried to get the ages for two SBs by using the UV-NIR colors,
but failed to determine them uniquely. 
This suggests that the SB ages cannot be determined from the UV-NIR SEDs alone.

The aim of this work is to investigate the SB galaxies studied by 
Gordon et al. (1997) with additional IRAS data and to derive the
accurate age and the dust attenuation of the SB population.
In \S 2, we briefly describe our model, 
in \S 3 we describe our data sample, and in \S 4 we
demonstrate that we are successful in disentangling the burst age
and the dust attenuation in the UV-FIR SEDs, and derive the burst
parameters for 22 individual SB galaxies.
In \S 5, we discuss cosmological implications of our
results and conclude this article in \S 6.

\section {Model of starburst galaxy}

Even in the SB galaxies, the burst population does not always
dominate the galaxy's light, to which the underlying population contributes
significantly, and sometimes exceeds the burst light if the SB is
already aged. We, therefore, adopt a realistic model of star-forming spirals
by superposing the burst population on the underlying galaxy. 
To model the SEDs of SB galaxies, both a stellar evolutionary
synthesis model and a dust radiative transfer (RT) model are needed.
We use the stellar evolutionary synthesis model of Kodama \& Arimoto
(1997), the latest version of Arimoto \& Yoshii (1986; 1987).
We calculate accurately the radiative transfer of the stellar photons
through the dust by solving the transfer equations 
(Takagi, Vansevi\v cius, \& Arimoto 1998). Main ideas concerning the
structure of SB galaxies are borrowed from Witt, Thronson \&
Capuano (1992) and Gordon et al. (1997).

\subsection {Structure of the starburst galaxy}

We have adopted a two-component model for the SB galaxy: 
the galaxy consists of the SB region and the underlying population
and is characterized 
with a free parameter $f_{SB}$, a mass ratio of the SB to the total galaxy.
The radiative transfer and the evolutionary
models are calculated for both components independently.
We assume that the densities of star 
and dust in the underlying population follow the modified
King's distribution:
\begin{equation}
\rho_{G}(r)=\frac{\rho_{G0}}{\left( 1+\left( \frac{\displaystyle r}
{\displaystyle r_{G}}\right)
^{2}\right) ^{%
\frac{1}{2}}},
\end{equation}
where $r_{G}$ means a core radius of the galaxy. The distribution of stars in
the SB region is modeled by a similar but steeper function:
\begin{equation}
\rho_{SB}(r)=\frac{\rho_{SB0}}{\left( 1+\left( \frac{\displaystyle r}
{\displaystyle r_{SB}}\right)
^{2}\right)
^{\frac{3}{2}}},
\end{equation}
where $r_{SB}$ means the core radius of the SB region. 
For the underlying and SB regions, the ratio of the total radius,
$R_{tot}$, to core radius, $r_{core}$, is fixed to be 
$\log (R_{tot}/r_{core})=2.2$. The distribution of dust in the SB is
assumed to be homogenous for the entire region. Both stellar and dust
distributions are integrated up to the same limiting radius.

As a test of validity of the geometry of our models, we have calculated a
few models following the suggestion by Gordon et al. (1997). They
assumed the SB
model as the dusty sphere with the inner dust-free cavity where the stellar
population is residing. No significant differences are found for
relatively small optical depths $\tau _{V}\lesssim 5$, 
where $\tau _{V}$ is integrated from the center to $R_{tot}$.

The model calculations are carried out for $f_{SB}=0.1-1.0.$ 
In this paper, the main attention is drawn to the two cases $f_{SB}=0.23$ and
$1.0$, which means $23\%$ and $100\%$ of galaxy mass are involved in
the starburst. The cases with $f_{SB} \le 0.2$ fail to cover 
the entire region of the
observed colors of the sampled galaxies in terms of age and $\tau_V$.
For the underlying, we adopt typical parameters for an Sc
galaxy, i.e., a mass of $10^{11}M_{\odot }$, and a radius of $10$ kpc.
If we assume 30\% SB mass to the underlying galaxy, it gives $f_{SB}=0.23$,
the SB radius $6.6$ kpc, and the mass $3 \times 10^{10}M_{\odot}$, respectively, 
if the the same average density is adopted in the SB region 
as in the underlying galaxy. The total mass of the 
galaxy is then $1.3\ 10^{11}M_{\odot }$. Such a size of the SB region is 
in agreement with the area covered by the diaphragm for the 
observations presented in Gordon et al. (1997).

The numerous tests with different geometrical parameters
reveal only small differences in the modeling results. However, 
in cases with $\tau_{V}>5$, the
precise simultaneous determination of the age and dustiness of SB is model
dependent, and have to be carefully calibrated.

\subsection {Radiative transfer in dusty galaxy}

The radiative transfer in the galaxy and SB is calculated using the RT code
developed by Takagi et al. (1998). The code is based on an assumption
of spherical symmetry with arbitrary radial distributions of stars and dust.
Direct integration along the rays is adopted. The steps of the mesh ($100$
shells) are determined automatically, depending on the stellar and dust
distributions, to fulfill the energy conservation. The ray tracing is done in
the way suggested by Band \& Grindlay (1985). The thermal emission from dust
is calculated by assuming an equilibrium of dust particles in the radiation
field and by taking into account self-absorption. Up to six isotropic 
scattering terms are calculated, and losses of energy due to the scattering 
do not exceed $0.3\%$. The total energy losses in the present models are 
less than $1\%$. The isotropic scattering is assumed for simplicity
without introducing significant errors, since we analyze optically thick 
cases.  We should note that the thermal 
emission, which is crucial for the present study, is independent of the 
adopted phase function. 

The results were tested by comparing with those obtained by Witt, Thronson, \& 
Capuano  (1992) and the output of RT benchmark by Ivezi\'{c} et al. (1997),
and were found to be consistent. Also no significant discrepancies are found
between our results and those obtained by DUSTY (Ivezi\'{c}, Nenkova
\& Elitzur 1997) for the model of a star enshrouded in a dusty envelope with a
central dust-free cavity.

For the present investigation, $\tau _{V}$ is adopted as a free parameter,
i.e., we mimic the amount of dust by changing the optical depth of our
models. That means the mass of dust depends on the details of model parameters.
This is why we do not derive directly the mass of dust but refer only 
$\tau _{V}$.
Any extinction law can be applied in the code, but for the
present investigation we limit ourselves with the MW and SMC extinction
curves (Pei 1992).  The albedo is taken from Pei (1992) for both 
of the extinction curves. We truncate the albedo curve for $\lambda > 7 \mu m$,
without introducing any significant error, in the calculation of the 
scattered light.

\subsection {Stellar evolutionary synthesis}

The evolutionary synthesis code developed by Kodama \& Arimoto (1997) 
is used for the 
modeling of the luminosity and chemical evolution of the galaxy and SB. The
code is based on the up to date ingredients,
such as stellar isochrones, spectral libraries, and chemical yields of various
elements, and has been successfully used in a numerous publications 
(e.g., Kodama et al. 1998; Kodama \& Arimoto 1998). 
The code gives the evolution
of synthesized spectra of galaxies (the wavelength range $0.09-3000 \mu m$)
in a consistent manner with the galaxy chemical evolution, i.e., 
the effects of 
stellar metallicity are explicitly considered (Arimoto \& Yoshii 1986). 

The underlying population of an Sc galaxy is modeled by following the 
prescription given by Arimoto, Yoshii \& Takahara (1992), i.e., 
taking the values of all free
parameters for an infall model; a SFR $k=0.190$ Gyr$^{-1}$, 
an infall rate $a=0.148$ Gyr$^{-1}$, Salpeter initial mass function 
(IMF) with a slope $x=1.35$, and lower and
upper stellar masses, $m_{\ell}=0.10M_{\odot}$ and $m_u=60M_{\odot}$,
respectively. We adopt $15$ Gyr as the age of galaxy
because the observed sample consists of nearby galaxies. 

The SB
population is modeled by assuming a closed box model with the SFR 
of $k_{SB}=10.0$ Gyr$^{-1}$. 
The SFR adopted is equivalent to that of a 
giant elliptical galaxy (Arimoto \& Yoshii 1987). 
The nebular continuum is neglected in the present models since we limit 
ourselves to ages $t > 10$ Myr when the amount of ionizing photons
$(\lambda \le 0.09 \mu m)$ is negligible except for the very beginning of the
bursts at $t \le 10$ Myr 
(Leitherer \& Heckman 1995; Fioc \& Rocca-Volmerange 1997).
The IMF for the SB is assumed to be Salpeter-like with $x=1.10$, and 
$m_{\ell}=0.10M_{\odot}$ and $m_u=60M_{\odot}$, respectively.
Such a flat IMF has been suggested for spheroidal
galaxies (e.g., Arimoto \& Yoshii 1987; Kodama \& Arimoto 1997)
and for the OB associations in the Milky Way 
(Massey, Johnson, \& DeGioia-Eastwood 1995).

The metallicities given by Calzetti, Kinney, \& Storchi-Bergmann (1994) for 
the SB galaxies are attributed to the ionized gas, and the averaged value 
for the sampled galaxies ($\log (O/H)+12=8.7 \pm 0.5$) 
is very close to solar.
Nevertheless, we calculate our models with the {\it initial} metallicity,
$Z_i=0.02$ and $0.008$, there is no significant differences.
Therefore, the {\it initial} metallicity of the SB region is adopted 
to be solar.

The SEDs of the underlying galaxy and the 
SB region at different ages are shown in
Fig.1. The thick solid line represents the underlying at $15$ Gyr. 
For the SBs, 
all SEDs are plotted for $\tau _{V}=1.0$. The optical depth for the underlying
galaxy is fixed at $\tau_{V}=0.5$ throughout the paper.

\section {Observational data}

The observational data of SB galaxies are taken from Gordon et al.
(1997). They presented the homogeneous sample 
of aperture-matched observations of $30$
SB galaxies from the far-UV to NIR. The size of the diaphragm of the 
observations is $10\times 20$ arcsec, 
which is equal to $\sim 4.5$ kpc at their sample's median distance of $60$ Mpc.

We have supplemented the NIR colors for $4$ galaxies (NGC1510, NGC5236, NGC5253
and NGC7552) by taking data from available multiaperture photometry (de
Vaucouleurs \& Longo 1988), which allows us to reduce the observed
magnitudes to the effective aperture used by Gordon et al.(1997).

The IRAS\ flux densities are taken from NED (NASA/IPAC Extragalactic Database)
and have been color corrected. Two galaxies 
(Mrk 309 and Mrk 542) in the sample have 
only upper limits for the $100\mu m$ flux density. It is important to note 
that the IRAS\ colors 
are derived for the total galaxy while our sample (UV-NIR)
is aperture limited. Nevertheless, taking into account the conclusion drawn by 
Calzetti et al.(1995) concerning the strong central concentration of FIR 
in the galaxies of the present sample, we believe that the necessary aperture 
correction does not change significantly our main conclusions.

Finally we end up with a sample complete in B, H, and IRAS $60$ and $100$
$\mu m$ fluxes for $22$ galaxies. The choice of present colors B, H, and FIR 
is done after an analysis of all the observational data given 
by Gordon et al.(1997).
{\it Only in the case of B-H and $L_{\rm FIR}/L_{\rm H}$, 
the disentangling of the age and dustiness is successful.}
$L_{\rm FIR}$ is the FIR luminosity 
derived from $60$ and $100$ $\mu m$ fluxes (Londsdale et al. 1985). 
Table 1 gives the observations and derived parameters of the SB galaxies: 
column (1) gives the name of galaxy, columns (2) and (3) indicate B-H and 
$\log (L_{\rm FIR}/L_{\rm H})$, respectively, and
columns (4)-(11) give the derived 
age in Myr and the optical depth (see \S 4) for both the MW and 
SMC extinction curves.

\section {Results}

The comparison of model results with the observed properties of
SB galaxies is divided into two
steps: a determination of the age and dustiness of SB galaxies, and
a comparison of the observed SEDs with model predictions for the age
and the opacity sequences.

\subsection {Simultaneous determination of the starburst age and
dustiness}

The calibration of SB galaxies with $f_{SB}=0.23$ 
by the age and dustiness for the MW extinction curve
is shown in Fig.2a.
The data points are 
taken from Table 1. The large circle indicates the underlying galaxy. 
An area indicated by a dashed curve represents the region 
occupied by Coma spirals 
(Gavazzi, Randone \& Branchini 1995). The solid lines show the 
isochrones with increasing $\tau _{V}$ from the left ($\tau_{V}=0.5$)
to the right ($\tau _{V}=10$). The 
dot-dashed lines show the isoopaques with
increasing age of the SB from the top ($t=10$ Myr) to the bottom ($t=1$ Gyr).

{\it It is obvious that FIR\ luminosity of galaxies helps to break the
age-dustiness degeneracy, and we have the unique solution for the SBs
with $\tau _{V}\leq 5$}. The range of $\tau _{V}$ validity depends on 
$f_{SB}$ assumed. Fig.2b shows the same diagram as Fig.2a but 
for $f_{SB}=1$,
i.e., the galaxy has 100\% SB population.
As we can see in Fig.2b, 
the applicable range of $\tau _{V}$ is significantly larger, but the shape
of the grid remains very similar to the case shown in Fig.2a.
Therefore, we expect that the assumed value of the free parameter 
$f_{SB}=0.23$ does not influence our conclusions significantly.
The derived ages and optical depths $\tau _{V}$ for individual SB
galaxies are given in Table 1 for $f_{SB}=0.23$ and $1.00$.
We note that the burst age and the opacity are nearly the 
same for different values of $f_{SB}$. The resulting age ranges from 
$t=10$ Myr to $t=500$ Myr, which agrees well
with Gordon et al.'s (1997) estimate for the present sample of SBs.

Figs.2c, d are the same as Figs.2a, b,
respectively, but for the SMC 
extinction curve. We note that the burst age and the opacity 
are nearly the same for different extinction laws.
{\it We therefore can conclude that the present method is robust and
does not depend on the details of extinction curve and the 
burst strength.}

\subsection {SEDs of dusty starburst galaxies}

The parameters given in Table 1 are used for grouping of the SB galaxies by
the age and the opacity. Two groups of galaxies are defined: (a) those with
approximately the same ages ($t=20-30$ Myr) but different
optical depths ($\tau_{V}=0.5-5.0$), 
and (b) those with similar optical depths ($\tau_{V}=0.9-1.5$) but different
ages ($t=10-400$ Myr). The SEDs inside these groups are averaged and
plotted in Figs.3a-h. The model SEDs are superposed on the
observational data without any fitting, but both are normalized at the 
H band to $1.0$. 

Figs.3a-d and Figs.3e-h represent the two different extinction
curves (MW and SMC, respectively). The SEDs in Figs.3a,b,e,f
represent $\tau_V$ sequences and in Figs.3c,d,g,h age sequences.
The influence of the extinction law is significant
in both the UV and FIR spectral regions. The behavior of the SED in
UV can be predicted from the shape of the extinction curve only in
the case of $f_{SB}=1.0$. A strong nonlinearity is observed for
$\tau_V>2$ in case of $f_{SB}=0.23$, where the stronger influence of $f_{SB}$ than
of the extinction law on the UV part of SED is observed, since 
the significant obscuration in the UV hides SB population
and the underlying galaxy, which is characterized by a MW extinction curve,
becomes dominant. FIR emission is affected more
regularly by the extinction curve and the burst strength, i.e., the FIR part
of SED becomes cooler for the SMC extinction law and brighter for $f_{SB}=1$.
As we can judge from the far-UV end of the SEDs, the extinction law is
steeper than MW but flatter than SMC. Therefore, Calzetti et al.
(1994) and Gordon et al. (1997) conclusions concerning the far-UV
shape of the extinction curve in SB galaxies are confirmed.

    It is important to note that the {\it observed} difference in the UV part
of SED is much smaller for the age sequence 
than for the $\tau_V$ sequence, but the trend is opposite in the FIR region. Along
the $\tau_V$ sequence, the UV-NIR part of SED becomes fainter, while
the FIR part becomes more luminous. On the other hand, both the
UV-NIR and FIR parts of SED becomes fainter along the age sequence
as the stellar population gets older. 

Figs.3a-h show that B,H
and FIR colors in general give a reasonable estimation for the
SEDs of SB galaxies from the UV to FIR.
Nevertheless, we 
have to stress that Figs.3a-h are given without any fitting of
the parameters, i.e., they represent a general behavior of our
models used to determine the age and $\tau_V$. Moreover, the observational
data are averaged for different age and dustiness groups.
The fine tuning of model parameters for each particular galaxy
is a subject for a forthcoming paper.

\section {Discussion}

Gordon et al. (1997) suggested that the dust distribution in
SB is best described by a clumpy media, which is
not adopted in our modeling. However, we should note that 
our Figs.2a-d would not change even if we take into account clumpy
media, because they can be scaled in B-H and 
$L_{\rm FIR}/L_{\rm H}$ independently of the dust
attenuation; thus our arguments are
safe so far as we do not discuss the mass of dust which is 
model dependent. Moreover, Witt \& Gordon (1996) suggested that the high 
conversion efficiency of 
UV/optical radiation into far-IR thermal dust emission 
observed in SB galaxies and mergers is due in large part to a change
in the structure of the interstellar medium in such systems toward a 
{\it more homogeneous widely extended distribution}. Taking into
account this suggestion and the curves corresponding to homogeneously 
extended distribution in their Fig.19 (the density ratio of 
interclump-to-clump media, $k2/k1 \geq 0.3$) , 
we can expect that our definition of 
$\tau_V$ for homogeneous distribution of dust is linearly related with
their $\tau_V$ calculated for the clumpy media.

Figs.3a-h clearly reveal two problems: a discrepancy of the model SEDs
and the observed ones in UV region, and
a difference of the predicted and observed fluxes in mid-infrared 
region (MIR).

The first point was discussed extensively by Gordon et al. (1997). We
have confirmed their conclusion that the extinction curve in the SB galaxies is
steeper than in MW. Therefore, from the comparison of the ages 
and $\tau_V$'s in Table 1, we conclude that the uncertainty of the 
extinction curve in the UV is not serious for the present study.
However, the deviation of the MW and SMC 
extinction curves adopted by us from the real one cannot significantly
change our results, because the main diagrams (Figs.2a-d) are based on the
bluest band B which is not influenced yet by the UV steepening of 
the extinction curve, and because the
supplementary amount of the energy absorbed in far-UV does not increase
significantly the FIR flux (Trewhella, Davies \& Disney 1997).

The second point, probably, can be explained by the 
shortcomings of our RT model, because the circumstellar dust and 
emission from very small grains are not included yet. 
The circumstellar dust mainly 
influences $12\mu m$ flux and is dependent on the number of stars with dusty 
shells in galaxy. 
The influence of small grains on the IRAS colors was discussed by 
Boulanger et al. (1988). They concluded that IRAS\ $12$ and $25\mu m$ colors 
depend on the non-equilibrium emission from small 
grains while $60$ and $100\mu m$
colors depend on the equilibrium emission from large grains. 
Extensive discussion 
on the applicability of so called ''classical'' grain model to fit the MIR 
($5-40\mu m$) fluxes from the interstellar medium was given by Helou, Ryter \& 
Soifer (1991). Therefore, our model represents 
correctly the FIR
emission ($60$ and $100\mu m$) from the grains in thermal equilibrium,
which is calculated explicitly by our RT code. Moreover, we are using 
only the FIR
colors for age-dustiness determination, and can neglect 
the discrepancies between
the predicted SEDs and the observed colors in the MIR region.

Figs.3a-h indicate that the observed FIR flux ratios $S_{60}/S_{100}$
(dust temperature indicator) are smaller than the model
predictions. This ratio depends on the dust temperature
and on the emissivity law of dust grains. The dust
temperature depends on the distribution of stellar density,
which is rather approximate in our model, a flatter one
would result in the cooler dust temperature. We assume the
emissivity with the power $\beta=2$ ($\lambda^{-\beta}$, Rowan-
Robinson, 1992) If $\beta<2$ is taken, the discrepancy of the
flux ratio would reduce (Hughes et al. 1997). However, such
a fine tuning of the model is beyond a scope of this article,
since both effects would not change significantly the total
FIR flux used for breaking the age-dustiness degeneracy.

We have demonstrated that the use of SEDs from far-UV to FIR is
quite effective in disentangling the age and opacity of the SB
galaxies. For deriving the opacity, Calzetti, Kinney, \& 
Storchi-Bergmann (1996) and 
Calzetti (1997) assumed that the intrinsic line ratios
are given by a simple model with the optically 
thick limit, the so-called Case B of Osterbrock (1989), and 
estimated the opacity from the ratios of attenuated
hydrogen emission lines. Although the opacities derived by two  
methods are correlated, the opacity given by
the line method is smaller than the FIR method.
This in turn should give systematically older ages for the line method.

One should keep in mind that some of the SB galaxies assigned for ages younger 
than 20 Myr could well be much younger if their colors are significantly 
reddened by possible nebular continuum (Fioc \& Rocca-Volmerange 1997).
However, even if this is the case, the probability for galaxies younger
than 10 Myr to reside in this region should be very small, because
of short timescale for that stage. Additionally, if galaxies are 
UV-selected as in Gordon et al. (1997), it would be almost impossible
to find such very young SB galaxies due to strong dust obscuration.

Before we conclude this article, we briefly discuss two cosmological
implications of the present results. First, if we define the primeval
galaxies as massive galaxies at the onset of the first episode of major 
star formation, presumably progenitors of the present-day ellipticals
at high redshifts, these galaxies should look like the SB galaxies 
in which burst population dominates the galactic light. 
Thus, it should be possible to determine the age
and optical depth of the primeval galaxy candidates with a help of
the rest-frame B-H {\it vs.} $L_{\rm FIR}/L_{\rm H}$ diagram. 
These galaxies should be spectacular in the rest-frame FIR regime. 
At the highest redshifts, the FIR $60$ and $100\mu m$
fluxes conveniently fall on into the region of submillimeter
atmospheric windows at $350$, $450$, $750$ and $850\mu m$.
Therefore, a search for the primeval galaxies with infrared satellites
and submillimeter telescopes, such as 
IRIS (Infrared Imaging Surveyor; Murakami et al. 1998) and 
LMSA (Large Millimeter Submillimeter Array; Ishiguro et al. 
1994), would be promising. Indeed, the sensitivity of submillimeter arrays is
expected to achieve the greatest advances in the immediate future 
(Hughes et al. 1997). 

Second, once the rest-frame B-H and $L_{\rm FIR}/L_{\rm H}$
are observed for high redshift galaxies including the Lyman-break galaxies 
(Steidel et al. 1996), radio galaxies (Dunlop et al. 1996), 
and QSO host galaxies, 
the ages and optical depths of these galaxies 
can be determined uniquely. Then, 
the resulting age should play the role of a cosmological clock that gives
the age-redshift relation of galaxies at high redshifts and will constrain 
the cosmological picture of the Universe. The resulting $\tau_{V}$
should in turn give a correction factor for the UV light, which will
make it possible to derive the correct SFR 
at high redshifts. In other words, it could be misleading to derive
the SFR for high redshift galaxies from the UV light alone.

\section {Conclusions}

 In an attempt to break the age-dustiness degeneracy of
the SEDs of starburst galaxies, we have built a new spectral
evolution model of galaxy that included both stellar and dust effects. 
The evolutionary code of Kodama \& Arimoto (1997) is used to model
the spectral and chemical evolution of galaxies, and the radiative transfer 
is calculated using the RT code developed by Takagi et al. (1998).
Applying the model to 22 nearby SB galaxies sampled by 
Gordon et al. (1997), we have shown that the unique solution can be
sought if the FIR luminosity is used for the starbursts with
$\tau_{V} \le 5$. Except for two opaque galaxies with $\tau_{V} > 5$,
the resulting SB ages and opacities are in the
range $10 \le t \ ({\rm Myr}) \le 500$ and $0.5 \le \tau_{V} \le 5.0$,
when the burst mass fraction $f_{SB}=0.23$ is assumed.
The result is robust and almost independent of 
the details of the dust distribution, the extinction curve, and the burst strength.
Even in a case of the SB dominant galaxies, the age and $\tau_{V}$ are 
uniquely assigned on the rest-frame B-H {\it vs.} 
$L_{\rm FIR}/L_{\rm H}$ diagram. 
For high redshift galaxies, these wavelength regions conveniently fall
on into the NIR, MIR, and submillimeter regions.
With the rapidly growing sensitivity of the submillimeter detectors, it
should soon become possible to determine the age and $\tau_{V}$ of 
star-forming galaxies at redshifts $z \simeq 3$ and beyond.
Accurate estimates of $\tau_{V}$ for the high redshift galaxies might
require a substantial revision of the broad picture of star
formation history over the Hubble time proposed by Madau et al. (1996).

\acknowledgments

We are very grateful to the referee, K.D.Gordon, for helping us to 
improve the paper significantly.
T.T. wish to thank to T.Shibazaki and R.Hoshi for their continuous
support and fruitful discussions.
This research has made use of the NASA/IPAC Extragalactic Database 
(NED) which is operated by the Jet Propulsion Laboratory, 
California Institute of Technology, under contract with the
National Aeronautics and Space Administration.
This work was financially supported in part by a Grant-in-Aid for 
the Scientific Research (No.0940311)
by the Japanese Ministry of Education, Culture, Sports and Science.

{}

\newpage
\begin{deluxetable}{lccccccccccc}
\tablenum{1}
\footnotesize
\tablewidth{0pt}
\tablecaption{Observed colors, derived ages and optical depths of SB galaxies}
\newcommand{\lw}[1]{\smash{\lower1.8ex\hbox{#1}}}
\newcommand{\lww}[1]{\smash{\lower0.3ex\hbox{#1}}}
\tablehead{
&&&
\multicolumn{4}{c}{MW}&&
\multicolumn{4}{c}{SMC}\\
\colhead{\lw{Galaxy}} & \colhead{\lw{B-H}} &
\colhead{\lw{$\log(\frac{L_{\rm FIR}}{L_{\rm H}})$}} &
\multicolumn{2}{c}{Age $[Myr]$}&
\multicolumn{2}{c}{$\tau_{V}$} &&
\multicolumn{2}{c}{Age $[Myr]$}&
\multicolumn{2}{c}{$\tau_{V}$}\\
\cline{4-7}
\cline{9-12} 
&&& {\scriptsize \lww{\hspace{1mm} $0.23^\dagger$}}
&{\scriptsize \lww{\hspace{1mm} $1.00^\dagger$}}
&{\scriptsize \lww{\hspace{1mm} $0.23^\dagger$}} 
&{\scriptsize \lww{\hspace{3mm} $1.00^\dagger$}}
&
&{\scriptsize \lww{\hspace{1mm} $0.23^\dagger$}}
&{\scriptsize \lww{\hspace{3mm} $1.00^\dagger$}}
&{\scriptsize \lww{\hspace{1mm} $0.23^\dagger$}}
&{\scriptsize \lww{\hspace{3mm} $1.00^\dagger$}}
\tablecomments{$\dagger$ The ratio of the SB mass to the  
total mass of galaxy, $f_{SB}$ \\
\hspace*{18mm}: The ages and optical depths derived by the upper limit
of $L_{\rm FIR}$.}
}
\startdata
NGC 1140 & 1.89 & 0.72 & 25 & 30 & 0.7 & 0.8 && 30 & 40 & 0.5 & 0.6  \nl
NGC 1510 & 2.74 & 0.22 & 400 & 500 & 1.1 & 0.9 && 400 & 500 & 1.1 & 0.8 \nl
NGC 1569 & 1.37 & 1.17 & \hspace{-2mm}{\tiny $<$}10 & 15 & 1.0 & 1.1 &&
\hspace{-2mm}{\tiny $<$}10 & 15 & 0.8 & 0.9 \nl
NGC 1614 & 3.97 & 1.14 & 30 & 25 & \hspace{-2mm}{\tiny $>$}$5.0$ & 4.3 &&
 30 & 30 & \hspace{-2mm}{\tiny $>$}5.0 & 4.0\nl
NGC 4194 & 3.29 & 1.01 & 25 & 30 & 3.6 & 2.9 && 30 & 30 & 3.3 & 2.6 \nl
NGC 4385 & 3.27 & 0.66 & 100 & 130 & 2.4 & 2.2 && 100 & 200 & 2.2 & 2.1 \nl
NGC 5236 & 2.93 & 1.03 & 20 & 25 & 2.9 & 2.5 && 25 & 30 & 2.6 & 2.2 \nl
NGC 5253 & 1.80 & 1.31 & \hspace{-2mm}{\tiny $<$}10 & 15 & 2.0 & 1.9 &&
\hspace{-2mm}{\tiny $<$}10 & 15 & 1.8 & 1.6 \nl
NGC 5860 & 2.79 & 0.55 & 130 & 200 & 1.5 & 1.5&& 200 & 300 & 1.4 & 1.4 \nl
NGC 6052 & 2.29 & 1.11 & 15 & 20 & 2.1 & 1.9 && 20 & 25 & 1.9 & 1.7 \nl
NGC 6090 & 3.22 & 1.02 & 25 & 30 & 3.6 & 2.9 && 25 & 30 & 3.3 & 2.6 \nl
NGC 6217 & 3.31 & 0.84 & 50 & 60 & 2.8 & 2.5 && 50 & 80 & 2.6 & 2.4 \nl
NGC 7250 & 2.09 & 0.92 & 20 & 25 & 1.3 & 1.3 && 25 & 30 & 1.1 & 1.1 \nl
NGC 7552 & 3.69 & 1.02 & 30 & 30 & 5.0 & 3.5 && 30 & 30 & 4.5 & 3.1 \nl
NGC 7673 & 2.17 & 0.94 & 20 & 25 & 1.4 & 1.4 && 25 & 30 & 1.2 & 1.2 \nl
NGC 7714 & 2.64 & 0.88 & 30 & 30 & 1.9 & 1.8 && 30 & 40 & 1.7 & 1.6 \nl
Mrk 309  & 2.76 & 1.43 &\hspace{-1mm}{\tiny $<$}10: & \hspace{1mm}15:
 & \hspace{-1mm}{\tiny $>$}5.0: & \hspace{1mm}3.5: &&
\hspace{-1mm}{\tiny $<$}10: & \hspace{1mm}15: 
&\hspace{-2mm}\hspace{1mm}{\tiny $>$}5.0: & \hspace{1mm}3.1: \nl
Mrk 357  & 2.01 & 0.91 & 20 & 25 & 1.2 & 1.2 && 25  & 30 & 1.0 & 1.0 \nl
Mrk 542  & 2.79 & 0.40 &\hspace{1mm}400: &\hspace{1mm}400: &\hspace{1mm}1.5:
 &\hspace{1mm}1.3: && \hspace{1mm}300: & \hspace{1mm}400: &\hspace{1mm}1.3:&
 \hspace{1mm}1.2: \nl
IC 214   & 4.18 & 0.92 & 100 & 50  &\hspace{-2mm}{\tiny $>$}$5.0$ & 4.0 &&
 100 & 80 &\hspace{-2mm}{\tiny $>$}5.0 & 3.6 \nl
IC 1586  & 2.49 & 0.59 & 80 & 100 & 1.1 & 1.2 && 100 & 100 & 1.0 & 1.0 \nl
Haro 15  & 2.10 & 0.74 & 30 & 30 & 0.9 & 1.0 && 30 & 40 & 0.7 & 0.8 \nl

\enddata
\end{deluxetable}

\begin{figure*}
%\begin{center}
\epsfxsize=13.0cm
\centerline{\epsfbox{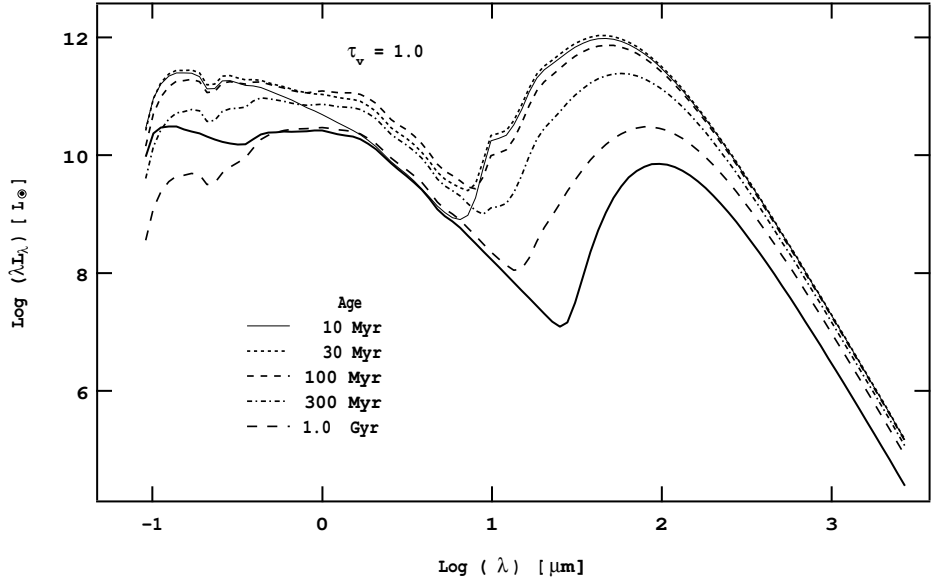}}
%\end{center}
\caption{The spectral energy distributions (SEDs) of SB region 
with a mass, $3 \times10^{10} M_{\odot }$ , $\tau _{V} =1.0$,
and different ages (thin lines) and of the 
underlying galaxy with a  mass,
$10^{11} M_{\odot }$, $\tau _{V} =0.5$, and $t=15$ Gyr (thick solid line).
The ages of SB region are shown in the figure. The MW extinction curve is applied.
\label{fig1}}
\end{figure*}

\begin{figure*}
%\begin{center}
%\epsfbox{file=fig2a,width=10.0cm}
%\end{center}
\epsfxsize=10.0cm
\centerline{\epsfbox{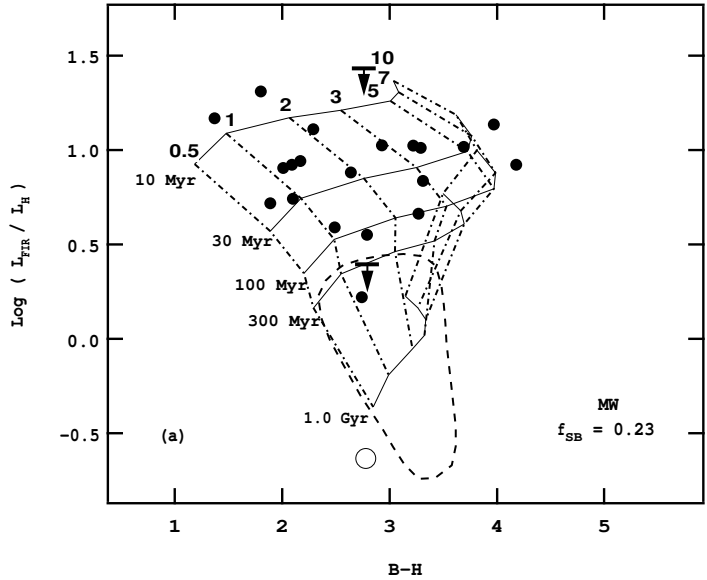}}
\caption{
(a) The calibration of SB galaxies by the age and dustiness.
All models are calculated with $f_{SB}=0.23$ and the MW extinction curve.
The filled circles are data points given 
in Table 1. The solid lines are isochrones with $\tau_V = 0.5,
1.0, 2.0, 3.0, 5.0, 7.0$, and $10.0$ from the left to the right.
The dot-dashed lines are isoopaques with $t = 10, 30, 100, 300 $Myr and
$1.0$ Gyr from the top to the bottom. 
\label{fig2a}}
\end{figure*}

\setcounter{figure}{1}
\begin{figure*}
%\begin{center}
%\epsfbox{file=fig2b,epsfx=10.0cm}
%\end{center}
\epsfxsize=10.0cm
\centerline{\epsfbox{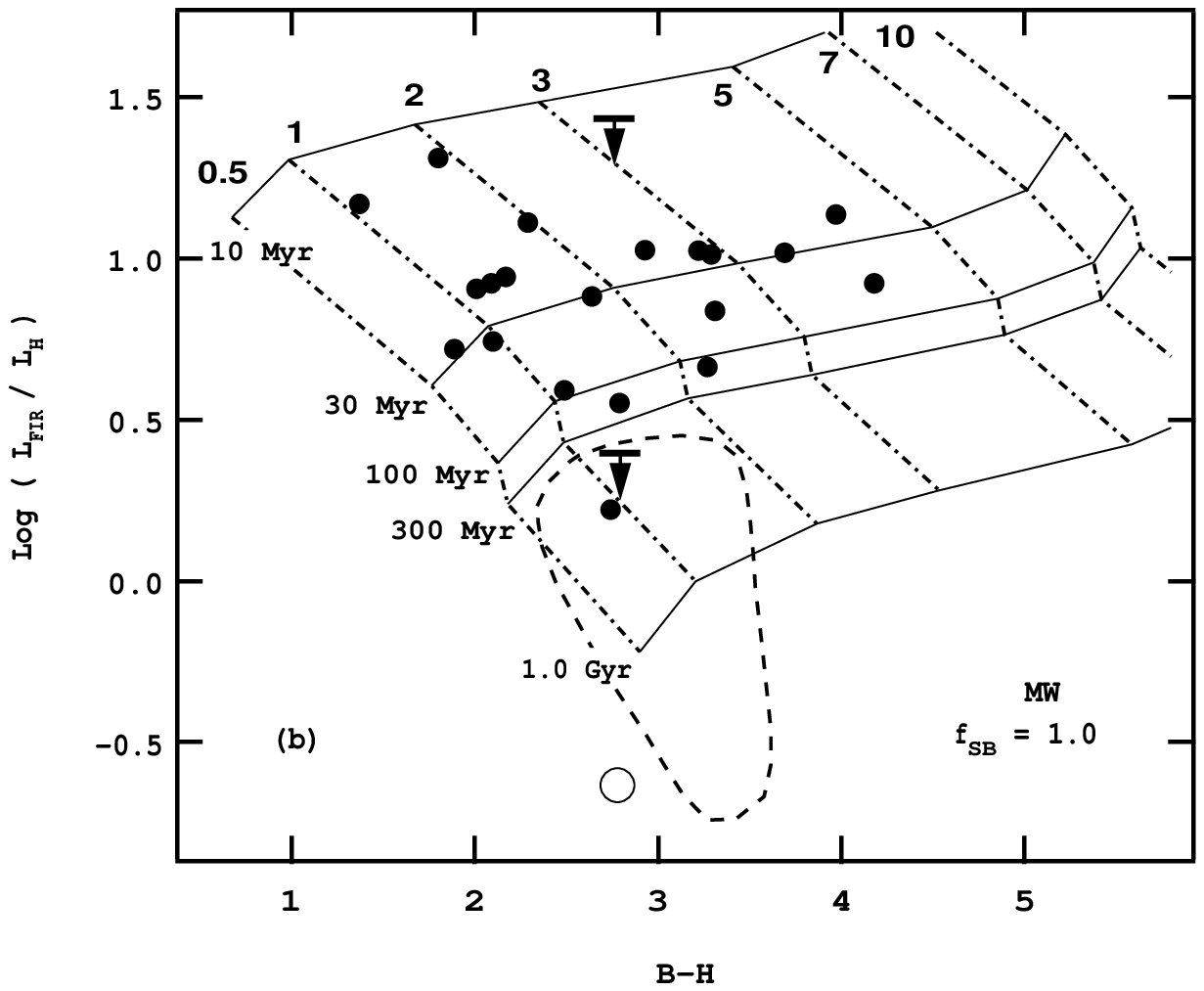}}
\caption{
(b) The same as (a) but for $f_{SB}=1.0$.
\label{fig2b}}
\end{figure*}

\setcounter{figure}{1}
\begin{figure*}
%\begin{center}
%\epsfbox{file=fig2c,epsfx=10.0cm}
%\end{center}
\epsfxsize=10.0cm
\centerline{\epsfbox{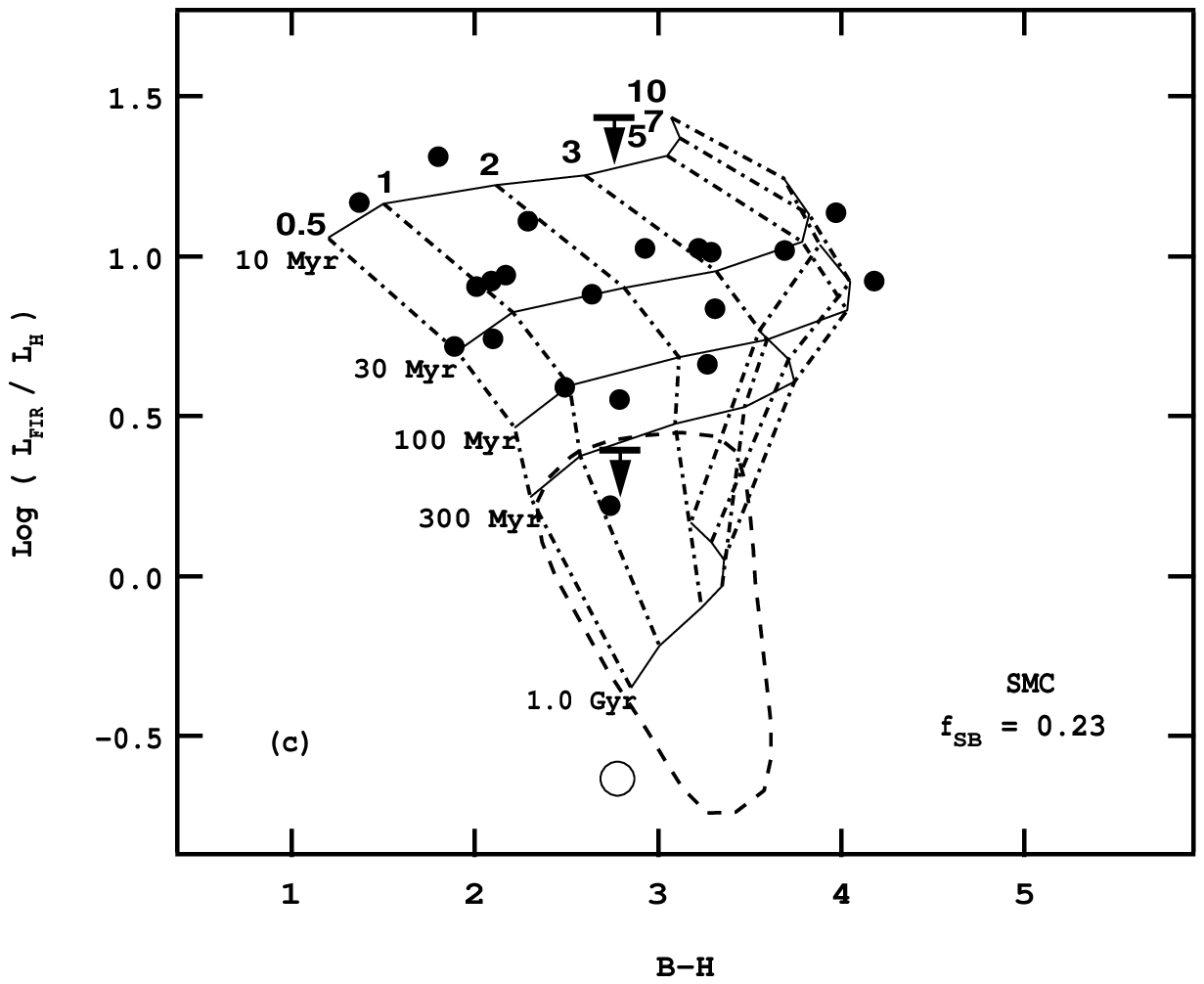}}
\caption{
(c) The same as (a) but with the SMC extinction curve.
\label{fig2c}}
\end{figure*}

\setcounter{figure}{1}
\begin{figure*}
%\begin{center}
%\epsfbox{file=fig2d,epsfx=10.0cm}
%\end{center}
\epsfxsize=10.0cm
\centerline{\epsfbox{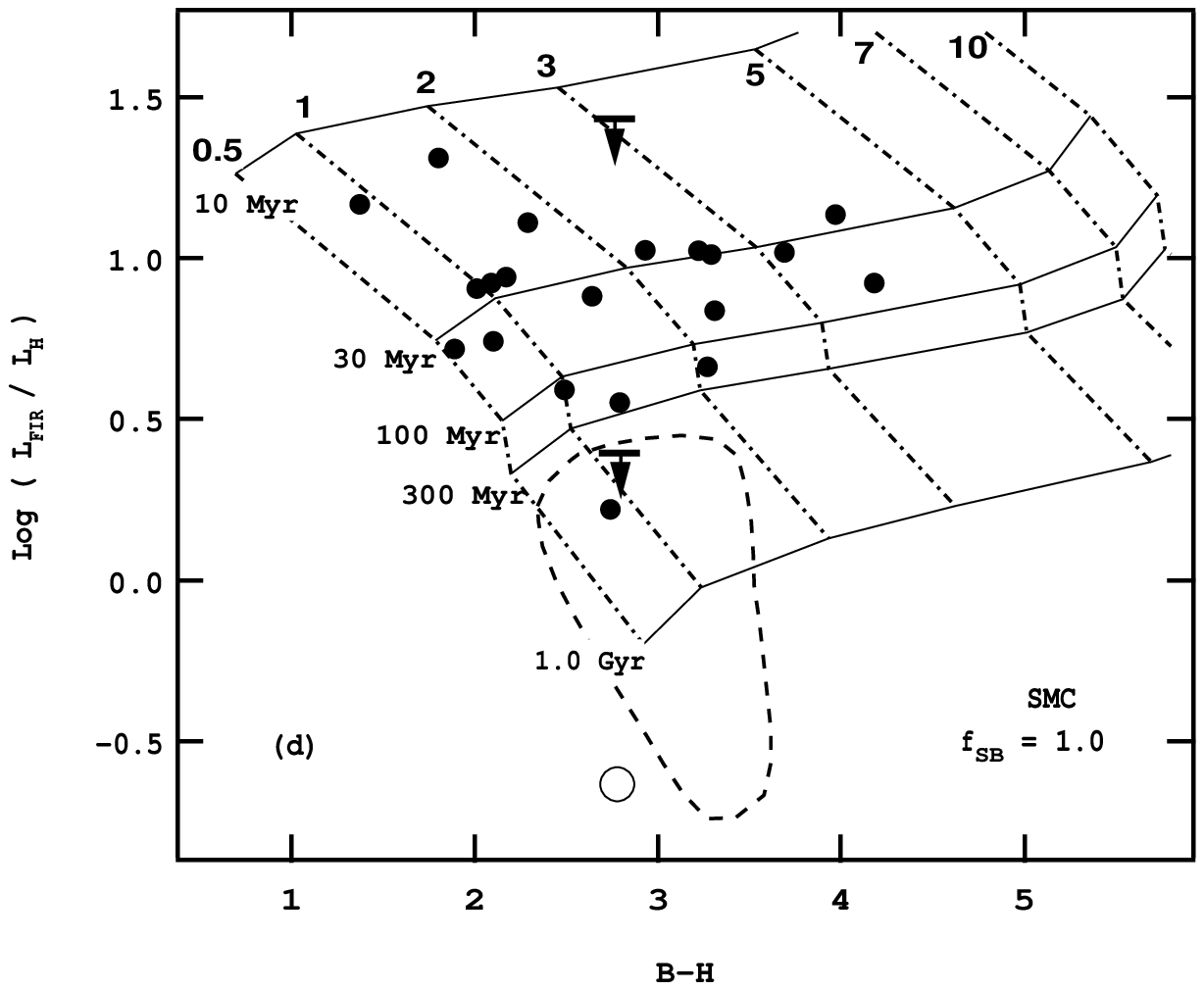}}
\caption{
(d) The same as (c) but for $f_{SB}=1.0$.
\label{fig2d}}
\end{figure*}

\begin{figure*}
%\begin{center}
\epsfxsize=13.0cm
\centerline{\epsfbox{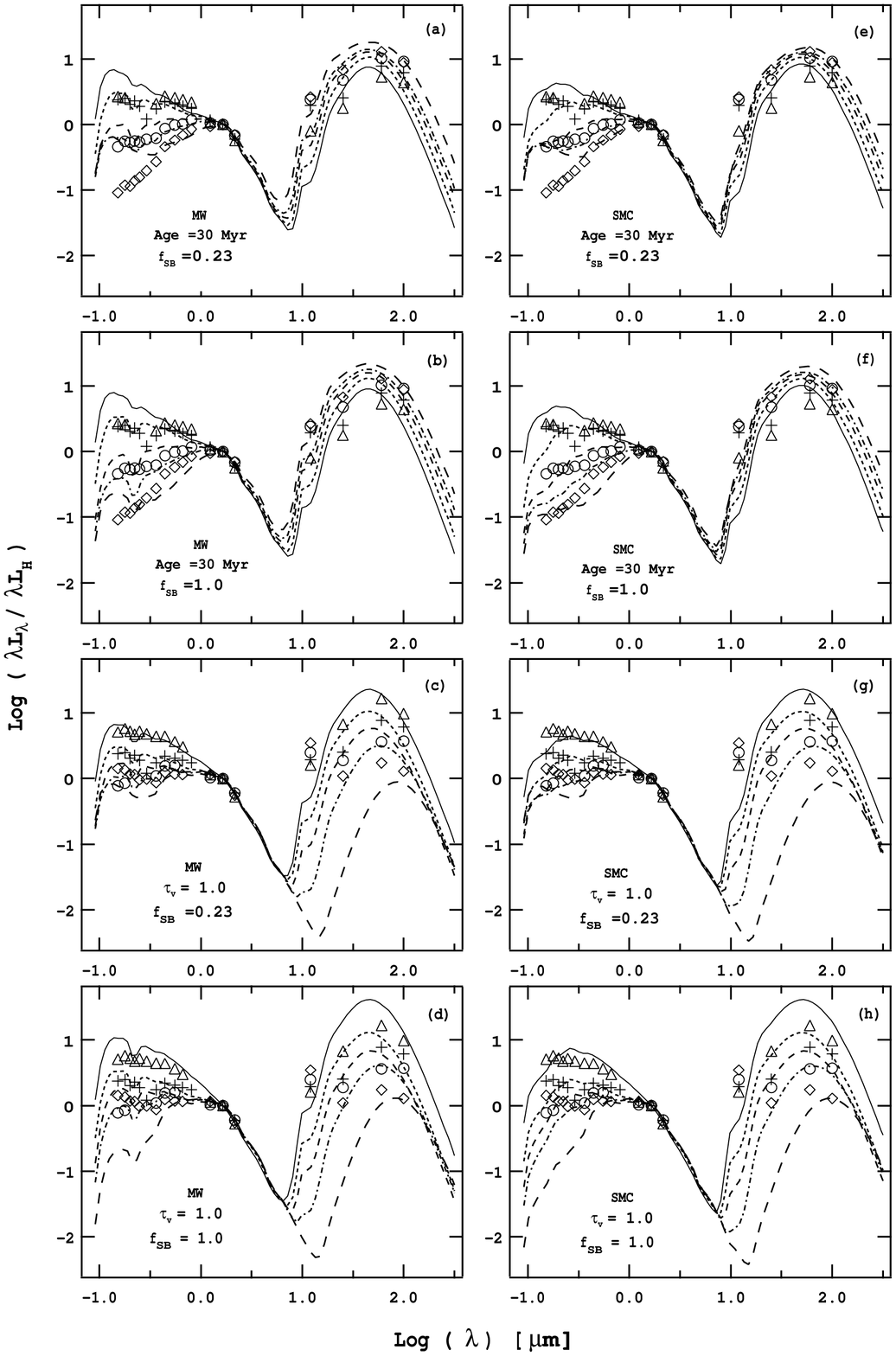}}
%\end{center}
%\caption{}
\end{figure*}

\setcounter{figure}{2}
\begin{figure*}
%\begin{center}
%\postscriptbox{3cm}{3cm}{}
%\end{center}
%\epsfbox{}
\caption{
Comparisons of the model SEDs with the averaged empirical SEDs within 
the groups of galaxies (see below for the definition of groups).
All SEDs are normalized at H band.
(a) The group of SB galaxies with similar ages 
$(t \simeq 20 \sim 30$ Myr) but different opacities. 
The galaxies are divided into four sub-groups by the 
optical depths, $\tau _V \simeq 0.5 $ (triangles), $1.0$(crosses),
$3.0$(circles), $5.0$(diamonds). The theoretical loci are
calculated with the age $t=30$ Myr, $f_{SB}=0.23$, the MW extinction
curve, and the optical depths $\tau_V=0.5$ (solid line),
1.0 (dotted line), 2.0 (dashed line), 3.0 (dot-dashed line), and
5.0 (long dashed line), respectively. 
(b) The same as (a) but for $f_{SB}=1.0$.
(c) The group of SB galaxies with similar optical 
depths $(\tau_{V} \simeq 0.8 \sim 1.5)$
but different ages, $t \simeq  10 $ Myr (triangles), $30$ Myr (crosses),
$100$ Myr (circles), and $400$ Myr (diamonds). 
The theoretical loci are calculated with $\tau_{V} =1.0$, $f_{SB}=0.23$,
the MW extinction curve, and the ages $10$ Myr (solid line), 
$30$ Myr (dotted line), $100$ Myr (dashed line),
$300$ Myr (dot-dashed line), and $1$ Gyr (long dashed line), respectively. 
(d) The same as (c) but for $f_{SB}=1.0$.
(e)-(h) The same as (a)-(d), respectively, but with the SMC extinction curve.
\label{fig3}
}
\end{figure*}


\begin{thebibliography}{}
\bibitem[]{} Arimoto, N., \& Yoshii, Y. 1986, A\&A, 164, 260
\bibitem[]{} Arimoto, N., \& Yoshii, Y. 1987, A\&A, 173, 23
\bibitem[]{} Arimoto, N., Yoshii, Y., \& Takahara, F. 1992, A\&A, 253, 21
\bibitem[]{} Band, D.L., \& Grindlay, J.E. 1985, ApJ, 298, 128
\bibitem[]{} Boulanger, F., Beichman, C., Desert, F.X., Helou, G., 
 Perault, M., \& Ryter, C. 1988, ApJ, 332, 328
\bibitem[]{} Calzetti, D. 1998, astro-ph/9806083
\bibitem[]{} Calzetti, D. 1997, AJ, 113, 162
\bibitem[]{} Calzetti,D., Bohlin, R.C., Kinney, A.L., Storchi-Bergmann, T.,
 \& Heckman, T.M. 1995,ApJ, 443, 136
\bibitem[]{} Calzetti,D., Kinney, A.L., Storchi-Bergmann, T. 1994,
 ApJ, 429, 582
\bibitem[]{} Calzetti,D., Kinney, A.L., Storchi-Bergmann, T. 1996,
 ApJ, 458, 132
\bibitem[]{} Connolly, A.J., Szalay, A.S., Dickinson, M., Subbarao,
 M.U.,  \& Brunner, R.J. 1997, ApJL, 486, L11
\bibitem[]{} de Vaucouleurs, A., \& Longo, G. 1988, 
 Catalogue of visual and infrared photometry of galaxies from 0.5
 micrometer to 10 micrometer (1961-1985), University of Texas Monographs
 in Astronomy, Austin: University of Texas
\bibitem[]{} Dunlop, J.S., Peacock, J., Spinrad, H., Dey, A.,
 Jimenez, R., Stern, D., \& Windhorst, R. 1996, Nature, 381, 581
\bibitem[]{} Fioc, M., \& Rocca-Volmerange, B. 1997, A\&A, 326, 950
\bibitem[]{} Gavazzi, G., Randone, I., \& Branchini, E. 1995, ApJ, 438, 590
\bibitem[]{} Gordon, K.D., Witt, A.N., Carruthers, G.R., Christensen, S.A., 
 \& Dohne, B.C. 1994, ApJ, 432, 641
\bibitem[]{} Gordon, K.D., Calzetti, D., Witt, A.N. 1997, ApJ, 487, 625
\bibitem[]{} Helou, G., Ryter, C., \& Soifer, B.T. 1991, ApJ, 376, 505
\bibitem[]{} Hughes, D.H., Dunlop, J.S., \& Rawlings, S. 1997, MNRAS,
 289, 766
\bibitem[]{} Ishiguro, M., Kawabe, R., Nakai, N., Morita, K.-I., 
 Okumura, S.K., \& Ohashi, N. 1994, in Astronomy with Millimeter and 
 Submillimeter Wave Interferometry, IAU Cool. 140, ASP Conf. Ser., 59, 405
\bibitem[]{} Ivezi\'{c}, \v {Z}., Groenewegen, M.A.T., Men'shchikov, A., 
 \& Szczerba, R. 1997, MNRAS, 291, 121
\bibitem[]{} Ivezi\'{c}, \v {Z}., Nenkova, M., \& Elitzur, M. 1997, 
 User Manual for DUSTY, University of Kentucky
\bibitem[]{} Kodama, T., \& Arimoto, N. 1997, A\&A, 320, 41
\bibitem[]{} Kodama, T., \& Arimoto, N. 1998, MNRAS, in press, astro-ph/9806029
\bibitem[]{} Kodama, T., Arimoto, N., Barger, A.J., \& 
 Arag\'on-Salamanca, A. 1998, A\&A, 334, 99
\bibitem[]{} Leitherer, C., \& Heckman, T.M. 1995, ApJS, 96, 9
\bibitem[]{} Londsdale, C.J., Helou, G., Good, J.C., Rice, W.
 1985, Cataloged Galaxies and Quasars Observed in the IRAS Survey,
 Jet Propuls. Lab., Pasadena, California
\bibitem[]{} Lilly, S.J., Le F\'evre, O., Hammer, F., \& Crampton, D.
 1996, ApJ, 460, L1
\bibitem[]{} Madau, P., Ferguson, H.C., Dickinson, M.E., 
 Giavalisco, M., Steidel, C.C., \& Fruchter, A. 1996, MNRAS, 283, 1388
\bibitem[]{} Massey, P., Johnson, K.E., \& DeGioia-Eastwood, K. 1995, 
 ApJ, 454, 151
\bibitem[]{} Meurer, G.R., Heckman, T.M., Lehnert, M.D., Leitherer, C.,
 \& Lowenthal, J. 1997, AJ, 114, 54
\bibitem[]{} Murakami, H., et al. 1998, ISAS Research Note No.651
\bibitem[]{} Osterbrock, D.E. 1989, Astrophysics of Gaseous Nebulae and Active
 Galactic Nuclei, University Science Books, New York
\bibitem[]{} Rowan-Robinson, M. 1992, MNRAS, 258, 787
\bibitem[]{} Rowan-Robinson, M., et al. 1997, MNRAS, 289, 490
\bibitem[]{} Pei, Y.C. 1992, ApJ, 395, 130
\bibitem[]{} Steidel, C.C., Giavalisco, M., Pettini, M., Dickinson, M.,
 \& Adelberger, K.L. 1996, ApJL, 462, L17
\bibitem[]{} Takagi, T., Vansevi\v cius, V., Arimoto, N. 1998, 
 to be submitted to PASJ.
\bibitem[]{} Trewhella, M., Davies, J.I., \& Disney, M.J. 1997, 
 Proc. Cold Dust Morphology,
 Conf. at WITS University in Johannesburg, South Africa, 22-26th January 1996
\bibitem[]{} Witt, A.N., Thronson, H.A., \& Capuano, J.M. 1992, ApJ, 393, 611
\bibitem[]{} Witt, A.N., \& Gordon, K.D. 1996, ApJ, 463, 681

\end{thebibliography}
\end{document}